\documentclass[aip,amsmath,amssymb,reprint,groupedaddress]{revtex4-1}

\usepackage{graphicx}
\usepackage{dcolumn}
\usepackage{bm}

\usepackage[utf8]{inputenc}
\usepackage[T1]{fontenc}
\usepackage{mathptmx}
\usepackage{etoolbox} 
\usepackage{hyperref} 
\usepackage{booktabs} 

\usepackage{xcolor}
\hypersetup{
    colorlinks,
    linkcolor={red!50!black},
    citecolor={green!50!black},
    urlcolor={blue!70!black} 
}

\makeatletter
\def\@email#1#2{%
 \endgroup
 \patchcmd{\titleblock@produce}
  {\frontmatter@RRAPformat}
  {\frontmatter@RRAPformat{\produce@RRAP{*#1\href{mailto:#2}{#2}}}\frontmatter@RRAPformat}
  {}{}
}%

\makeatother

\begin{document}
 \begin{flushright} \includegraphics[width=4cm]{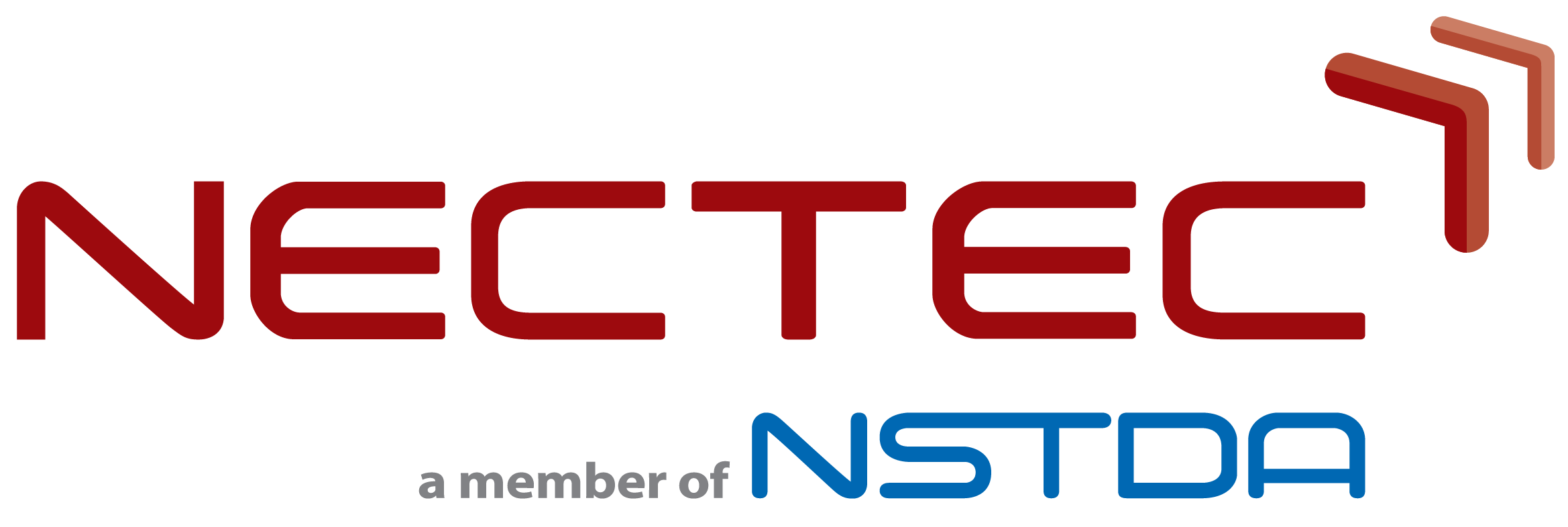} \end{flushright} 
 
\vspace{-4mm}
\title{Random number generation from a quantum tunneling diode}
\author{Kanin Aungskunsiri$^{*,\dagger}$} 


\author{Ratthasart Amarit$^\dagger$}
\author{Kruawan Wongpanya}
\author{Sakdinan Jantarachote}
\author{Wittawat Yamwong}
\author{Siriporn Saiburee}
\author{Sataporn Chanhorm}
\author{Apichart Intarapanich}
\author{Sarun Sumriddetchkajorn}


\affiliation{ \vspace{2mm}
National Electronics and Computer Technology Center,\\ 
National Science and Technology Development Agency,\\
Ministry of Higher Education, Science, Research, and Innovation,\\
\mbox{112 Thailand Science Park, Phahonyothin Road, Khlong Nueng, Khlong Luang, Pathum Thani, 12120 THAILAND}}

\date{\today , \href{https://doi.org/10.48550/arXiv.2002.02032}{arXiv.2002.02032}}


\begin{abstract}
\small
\vspace{-3mm} \noindent $^*$The author to whom correspondence may be addressed: \textit{kanin.aungskunsiri@nectec.or.th}. 

\vspace{1mm}\noindent $^\dagger$These two authors contributed equally to this work.\vspace{4mm}
\normalsize

\noindent\vspace{1mm}\textbf{ABSTRACT} \vspace{2mm}\\
\noindent Random numbers are important in many activities, including communication, encryption, science, gambling, finance, and decision-making. There is a strong demand for a hardware random number generator that could support cryptographic applications. In this work, we propose a quantum tunneling diode as a source of true randomness achieved by applying electrical current sweeps through the device and then harnessing a time-counting unit to measure fluctuation of current flows. Our approach can be implemented with inexpensive electronics and could be integrated into systems that require random numbers such as portable communication devices.

\vspace{2mm}\noindent \small
\textbf{Note:} 
\textit{This article may be downloaded for personal use only. Any other use requires prior permission of the author and AIP Publishing. This article appeared in \href{https://doi.org/10.1063/5.0055955}{Appl. Phys. Lett. \textbf{119}, 074002 (2021)} and may be found at \href{https://doi.org/10.1063/5.0055955}{https://doi.org/10.1063/5.0055955}.}

\normalsize


\end{abstract}

\maketitle




Random numbers have many uses in communication, science, mathematics\cite{pi}, testing, finance, and gambling. In communication, random numbers are critical for the encryption of sensitive information. The security of communication, in a practical sense, relies on the complexity of an encryption function; achieving that complexity, in turn, demands unpredictability of secret keys. In addition to quantum communication, real-world implementation of quantum-key-distribution protocol\cite{Bennett, Ekert, Townsend, Lo, Zhao, Zhang} demands true randomness in the process of establishing secret keys between trusted nodes—namely, both the sender and the receiver must randomly select quantum states for transmission and detection respectively. In stochastic simulation, random seeds are used in a mechanism to produce parameter values for simulation models with a variety of probability distributions\cite{ Crnjac}. Randomization is also key in gambling\cite{Johnston} and digital lotteries, especially in casinos, where a dedicated random number generator is vital. In the stock market, randomness is involved in a process of stocks distribution to buyers. Furthermore, verification via online transaction requires unpredictably random seeds for the generation of one-time passwords.

Random numbers can be obtained from various processes such as mechanical processes, computer programing\cite{Ecuyer}, chaotic process\cite{Gleeson, Uchida, Reidler, Argyris, Wishon}, radioactive decay\cite{Park}, atmospheric noise\cite{Marangon},  electronic noise\cite{Petrie}, and quantum phenomena. However, random number generation via mechanical processes and devices (e.g., coin flipping, Galton boards\cite{Galton}, and lottery wheels) have some intrinsic biases such that outputs are not equally distributed, which results in unreliable randomization. Computer programing generates pseudo-random numbers at a very fast speed in practice, but this source poses a security issue since an output value is calculated from a deterministic process. Electronic noise appears as a random disturbance in an electrical signal. It can serve a useful purpose for random number generation, but this source consists of intrinsic biases given that it is difficult to derive raw random outputs with good statistical properties. Atmospheric noise relates to a chaotic process that involves complexities cooperating with unknown parameters. Though atmospheric noise may be adequate in some applications, it can pose security risks. Namely, since the random signals are not isolated in a closed system, a third party could set up a device to obtain signals coming from the same origin. Exploitations of a quantum property of light\cite{Oberreiter} in various schemes, such as optical shot noise\cite{Rarity, Stefanov, Jennewein}, phase fluctuation of a laser\cite{Nie2015, Xu}, vacuum fluctuation\cite{Shi, Symul}, and photon arrival time measurements\cite{Nie2014, Wahl}, are approaches to obtain high bitrate quantum random number generation. 

\begin{figure*}[!htbp]
  \includegraphics[width=13cm,keepaspectratio]{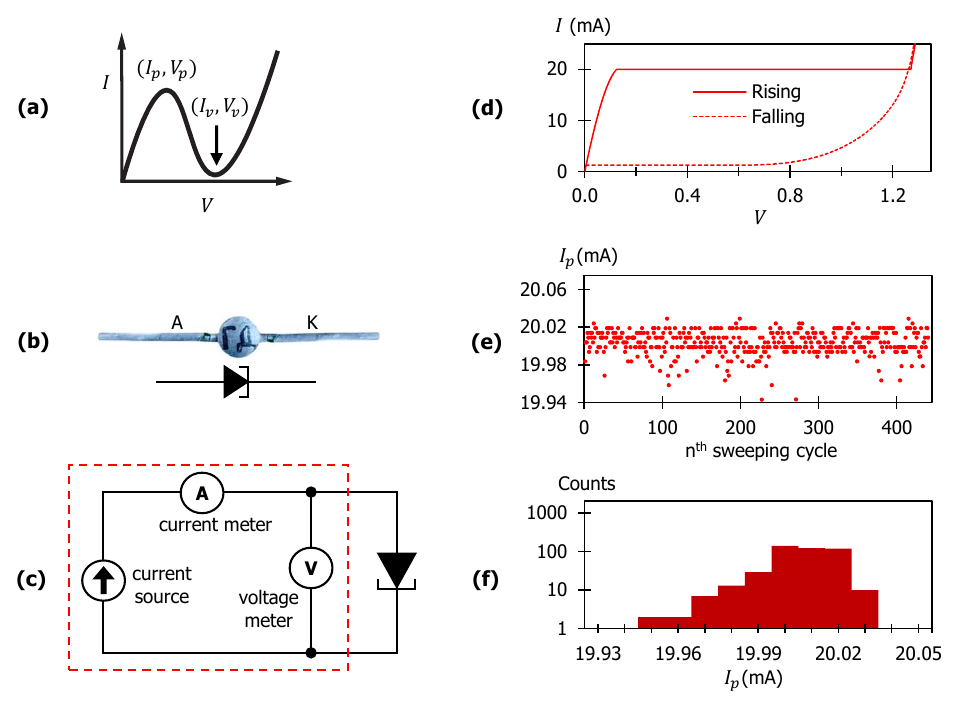}
  \vspace{-2mm}
    \caption{\textbf{(a)} Ideal I-V characteristic of a tunnel diode. \textbf{(b)} Tunnel diode used in this work (P/N: 3I201D). \textbf{(c)} Circuit diagram of the experiments. \textbf{(d)} I-V hysteresis loop measured from single current sweeping (round trip) according to (c). \textbf{(e, f)} Experimental results from 450 current sweeps showing (e) extracted values of $I_p$ and (f) distribution of $I_p$. In (c), equipment within the enclosed dashed line can be replaced with a current-source meter. In (d, e, f), the datasets were taken using a dedicated current-source meter (model: Keithley-2450).}
\end{figure*}

Regarding current technologies mentioned above, entropy sources acquired from the quantum origins are still expensive. Other entropy sources relate to the quantum tunneling effect in various semi-conductor devices such as resonant tunneling diodes\cite{Bernardo}, tunnel field-effect transistors\cite{Vezeteu}, and magnetic tunnel junctions\cite{Fukushima, Won, Vodenicarevic, Qin}. Among this device category, the tunnel diode as well as its exploitation for high quality random number generation would be of interest.

In this work, we propose an application of a tunnel diode\cite{Esaki} as a key component for obtaining an entropy source\cite{Aungskunsiri, ictp}. This device operates at room temperature and does not require expensive or complicated systems for its practical implementation.




A tunnel diode\cite{Esaki} is a solid-state device and one of many types of semi-conductor diodes incorporated with a p-n junction. See the supplementary material for a basic introduction of a tunnel diode. The device itself can operate in two modes: (i) a tunneling mode and (ii) a typical p-n junction mode. An ideal plot of current as a function of biased voltage (I-V characteristic) of a tunnel diode is illustrated in Figure 1a. As the figure shows, a tunnel diode has an N-shaped I-V characteristic such that there is a peak current ($I_p$) at a specific bias voltage ($V_p$). This characteristic is due to the collective behavior of tunneling electrons that results in the current flow. In fact, this behavior is different from the case of a single particle described by the quantum mechanics.

Regarding the N-shape of the I-V characteristic, sweeping current ($I$) and measuring voltage ($V$) across a tunnel diode results in a hysteresis cycle. We examined an off-the-shelf tunnel diode (P/N: 3I201D), as shown in Figure 1b, with a circuit diagram, as shown in Figure 1c. An I-V hysteresis loop obtained from our measurements is presented in Figure 1d. Here, $I_p$ is associated with the value of current where the voltage jumps to the other branch of the hysteresis loop.

Bernardo-Gavito et al.\cite{Bernardo} conducted a study of the hysteresis behavior associated with the generation of random numbers, examining the I-V characteristic by performing multiple current sweeps on a resonant-tunneling diode (RTD). In the prior-art, each forward current sweeping through the RTD took a different path along the I-V curve in a random manner. The values of $I_p$ as measured in their work fluctuated with a random distribution. We adopted this finding and looked for this behavior on a tunnel diode.

\begin{figure}[!b]
\includegraphics[width=6.0cm,keepaspectratio]{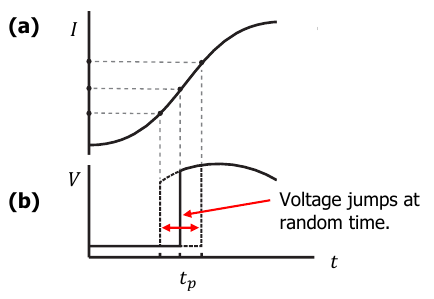}
\vspace{-2mm}
\caption{Method for measuring the value of $I_p$. A non-negatively sinusoidal waveform, having an amplitude far above the mean value of $I_p$, is applied for current sweeping through a tunnel-diode circuit. Time-series plots of applied current, as in \textbf{(a)}, and measured voltage across a tunnel diode, as in \textbf{(b)}, are illustrated together. Here, only the first part of the time-series plots is considered. At $t = t_p$, voltage jumps instantly corresponding to the time that the sweeping current reaches $I_p$.}
\end{figure}

\begin{figure*}[!htbp]
  \includegraphics[width=15cm,keepaspectratio]{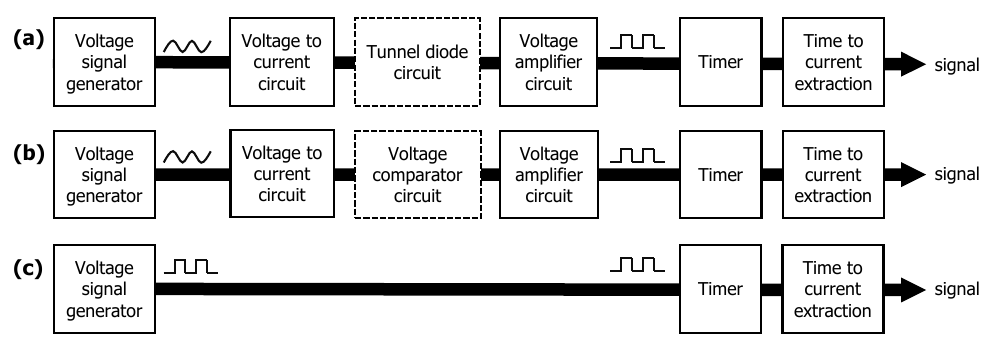}
  \vspace{-1mm}
  \caption{Flow diagrams of the experiment. We applied a voltage waveform to the electronic circuit and measured the response time for different scenarios: \textbf{(a)} with a tunnel diode circuit, and \textbf{(b, c)} without a tunnel diode circuit. In the beginning, a waveform generator (model: Hantek HDG2022B) was configured to produce non-negatively sinusoidal (a and b) and square (c) voltage with 100 Hz repetition rate. At the end of the flow diagrams, response time was converted to current. In (a), the voltage source was converted to the current source for sweeping the current in non-negatively sinusoidal waveform across the tunnel diode. A timer was implemented to obtain the values of $t_p$, which subsequently were converted to the values of $I_p$. In (b, c), the tunnel diode was removed from the circuit to check if any bias signals were originating from other circuit components apart from the tunnel diode itself.}
\end{figure*}

We applied a current-source meter to sweep current through a tunnel diode circuit and collected a dataset of I-V values from 450 current sweeps. The extracted values of $I_p$ are presented in Figure 1e and Figure 1f. Here, we obtained fluctuation of the values of $I_p$, which behaves in the same manner as found on the RTD\cite{Bernardo}. This outcome suggests the feasibility of harnessing a tunnel diode for random number generation. Unfortunately, a current-source meter is expensive. In addition, datasets acquired from the equipment are discrete-time signals. Detecting the values of $I_p$ that vary in the scale of microampere would be a highly time-consuming form of data acquisition.

Bernardo-Gavito et al.\cite{Bernardo} demonstrated another method to generate random numbers from the RTD that could be applied with the tunnel diode. Namely, a pulse train of current with constant amplitude was applied for current sweeping. By implementing a proper amplitude, the system generates uniform probability distribution of binary outputs. To sustain uniformity of the outputs, this approach would require a feedback control for real-time adjustment of the pulse-train amplitude.


Alternatively, we propose a simpler method that does not require a feedback control or any complicated system. The values of $I_p$ can be measured by monitoring the change of voltage. In detail, non-negatively sinusoidal waves described as a function of time with an amplitude $A$ and a frequency $\omega$, $I = A\sin (\omega t) + A$, are applied for current sweeping. We set the waveform's amplitude beyond all possible values of $I_p$ as illustrated in Figure 2, so we can be confident that the voltage always jumps to the other branch of the hysteresis loop in every sweeping cycle. Therefore, we can obtain the values of $I_p$ by measuring the values of the current signals at the time a sudden jump of the voltage occurs. As shown in Figure 2, we set $t = 0$ at the beginning of each current sweep. At $t = t_p$, the voltage jumps suddenly. Here, the measured values of $t_p$ are associated with the values of $I_p$.

\begin{figure}[!b]   %
  \includegraphics[width=8.0cm,keepaspectratio]{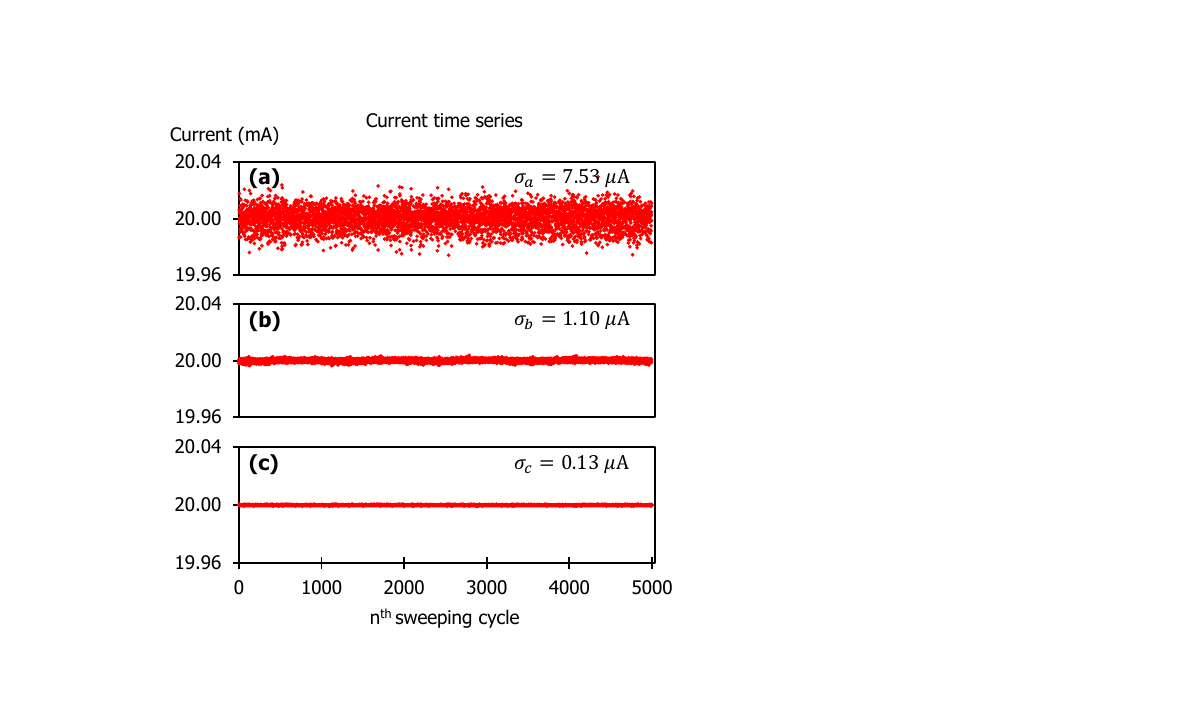}
  \vspace{-5mm}
    \caption{Current time series obtained from \textbf{(a)} a complete circuit having a tunnel diode circuit, \textbf{(b)} a voltage comparator circuit, and \textbf{(c)} a voltage signal generator, according to circuit diagrams (a)–(c) in Figure 3, respectively.}
\end{figure}

Regarding current technologies, we utilized a commercial time-counting unit as a solution for measuring $t_p$. Namely, a TEENSY$\textsuperscript{\textregistered}$ Development Board 4.0, having a retail price of \$19.95, is the generic time-counting module that we implemented in our prototype. This module has a time-resolution of about 6.7 ns (corresponding to 150 MHz of the sampling rate). The implementation of our method is shown in a flow diagram in Figure 3a. Obtaining the values of $t_p$ allows the values of $I_p$  to be realized by substituting the values of $t_p$ into the relationship $I_p=A\sin(\omega t_p) +A$, as described above.

Two additional measurements were designed to check whether noise was originating from other components. Firstly, we replaced the tunnel diode circuit from the system with a voltage comparator circuit, as shown in Figure 3b, to examine how much the tunnel diode contributed to the fluctuation. Secondly, we checked for bias signals that could come from the voltage signal generator by implementing the circuit diagram as depicted in Figure 3c. Data were collected from 5,000 current sweeps for each test. The results are presented in Figure 4.



Figure 4 presents experimental data on the amplitude fluctuation of the current signals measured from three different tests. We calculated the standard deviations ($\sigma$) of the datasets and realized that $\sigma_a > \sigma_b > \sigma_c$. Even though we detected noise from other parts of the circuit, the noise is very limited and could be neglectable. The noise might be coming from errors of the experiment; we leave this potential issue for further investigation.

\begin{figure}[!htbp]   %
  \includegraphics[width=8cm,keepaspectratio]{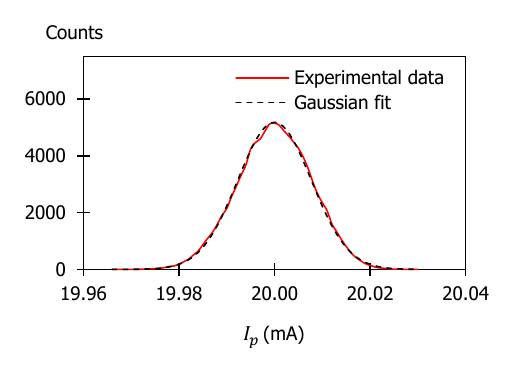}
  \vspace{-2mm}
  \caption{Histogram showing values of $I_p$ obtained from the tunnel-diode circuit (red solid line) in comparison with a Gaussian model plot (dashed line) featuring the standard deviation of 7.7 $\mu$A. The dataset is collected from 100,000 current sweeps.}
\end{figure}



By following the flow diagram in Figure 3a, we collected an additional dataset from 100,000 current sweeps and obtained the statistical distribution shown in Figure 5. We incorporated this result that suggests a Gaussian model for determining a suitable amplitude used for the current sweeping. The amplitude which is not too large allows us to obtain signals with good resolution. Accordingly, we managed to collect 20,000 samples per second.

In the implementation, raw signals were collected from the values of $t_p$. An offset is removed from the raw signals for digitization; consequently, binary outputs, as raw data, were extracted from the entropy source. Next, we followed the NIST SP800-90B\cite{Turan} and implemented a software\cite{software} provided by the NIST to assess the entropy source. This toolkit consists of two main steps: (i) determining the evaluation track, IID (independent and identically distributed) or non-IID, and (ii) estimating the minimum entropy rate of the raw data depending on the determined track. If the raw data fall into an IID track, entropy is considered from the most common value (MCV) estimation. Otherwise, entropy is assessed from the lowest value of entropy evaluated from 10 estimators, 
which are (1) MCV estimation, (2) collision estimation, (3) Markov estimation, (4) compression estimation, (5) t-Tuple estimation, (6) longest repeated substring (LRS) estimation, (7) multi most common in window prediction (multiMCW) estimation, (8) lag prediction estimation, (9) multiMMC Prediction estimation, and (10) LZ78Y prediction estimation.


\begin{table*}[!hbtp]
\footnotesize 
\centering 
\caption{Entropy assessment of raw data according to the NIST SP800-90B. Each dataset contains 40 million-bit sequences. The values of assessed min-entropy were selected from the lowest values from 10 estimators.}

\begin{ruledtabular}
\begin{tabular}{lccccc}
 & \multicolumn{5}{c}{\textbf{Min-entropy}}\\ \cmidrule{2-6} %
\multicolumn{1}{l}{\textbf{Estimator}} & \textbf{Dataset 1} & \textbf{Dataset 2} & \textbf{Dataset 3} & \textbf{Dataset 4} & \textbf{Dataset 5} \\ \cmidrule{1-6} 
1. MCV & 0.999 313  & 0.999 373  & 0.999 259  & 0.999 325  & 0.999 403  \\
2. Collision  & 1.000 000  & 1.000 000  & 1.00 0000  & 1.000 000  & 1.000 000  \\
3. Markov & 0.901 064  & 0.899 539  & 0.899 158  & 0.898 290  & 0.895 731  \\
4. Compression       & 0.619 087  & 0.622 348  & 0.623 679  & 0.624 235  & 0.625 367  \\
5. t-Tuple  & 0.872 663  & 0.867 644  & 0.874 429  & 0.867 644  & 0.872 663  \\
6. LRS    & 0.895 813  & 0.914673  & 0.852 600  & 0.881 406  & 0.865 663  \\
7. MultiMCW   & 1.000 000  & 1.000 000  & 1.000 000  & 1.000 000  & 1.000 000  \\
8. Lag Prediction    & 0.977 310  & 0.878 794  & 0.964 168  & 0.975 876  & 0.976 249  \\
9. MultiMMC Prediction  & 0.853 159  & 0.855 779  & 0.857 333  & 0.855 688  & 0.855 221  \\
10. LZ78Y Prediction  & 0.899 739  & 0.898 203  & 0.897 819  & 0.896 944  & 0.894 367  \\  \cmidrule{1-6}
\textbf{Assessed min-entropy}  & \textbf{0.619 087}  & \textbf{0.622 348}  & \textbf{0.623 679}  & \textbf{0.624 235}  & \textbf{0.625 367} 
\end{tabular}
\end{ruledtabular}
\end{table*}


According to the NIST SP800-90B\cite{Turan}, our raw data are determined as non-IID. Table I shows the results of min-entropy evaluated from 10 estimators. Five datasets of 40 million-bit sequences were examined. Here, the compression estimation provides the lowest value of min-entropy, given that the assessed min-entropy, $H_{\infty}$, was 0.62 on average. This value reflects the lower bound of random bits we can achieve from the raw data. As shown in Table I, the raw data have stable entropy.

These raw data were not ready for real-world use, and they did not pass the NIST SP800-22\cite{Rukhin} examination for statistical randomness. Post processing was mandatory to improve the raw data quality. We implemented Toeplitz-hashing extractor\cite{Ma, Zheng} with the raw data to render uniformly distributed random outputs. This extractor employs a binary matrix with a dimension of $m \times n$, where $m$ and $n$ correspond to the output and input bit-string length, respectively. $m$ is considered from the value of assessed min-entropy such that $m \leq nH_\infty$. Security parameter is a measure of the statical distance between the distribution of the dataset and the ideal uniform distribution. By implementing the Toeplitz extractor, the security parameter\cite{Ma, Zheng} of the post-processed data is estimated from $2^{(m-nH_\infty)/2}$.

In the implementation, we set $n=4096$. Hence, $m \leq 2539$. We chose $m=2296$. Thus the Toeplitz matrix has a size of $2296 \times 4096$, providing that the security parameter is below $2^{-100}$.

The post-processed data, as the final outputs, were re-checked with the NIST SP800-90B\cite{Turan}. The outputs at this stage passed the IID test. Therefore, the min-entropy was determined from the MCV estimation. Table II shows min-entropy assessed from the raw data and the final outputs. The result shows improvement of the statistical properties, and almost full entropy is achieved.


\begin{table}[!htbp]
\footnotesize
\centering
\caption{Entropy assessment of raw data versus final outputs according to the NIST SP800-90B. Final outputs in each dataset were extracted from 40 million-bit sequences of raw data.}
\begin{ruledtabular}
\begin{tabular}{lccccc}
 & \multicolumn{5}{c}{\textbf{Assessed min-entropy}} \\ \cmidrule{2-6}  
\multicolumn{1}{c}{\textbf{}} & \textbf{Dataset 1} & \textbf{Dataset 2} & \textbf{Dataset 3} & \textbf{Dataset 4} & \textbf{Dataset 5} \\
Raw data & 0.619 087  & 0.622 348  & 0.623 679  & 0.624 235  & 0.625 367  \\
Final outputs & 0.998 865  & 0.998 956  & 0.998 919  & 0.998 804  & 0.999 194  \\ 

\end{tabular}
\end{ruledtabular}
\end{table}


To validate the statistical randomness of the final outputs, a hundred datasets of 1,000,000-bit sequences were examined with 15 sub-methods in accordance to the NIST SP800-22 Test Suite Compliance\cite{Rukhin}. The test results, as presented in Table III, indicate that the final outputs passed the NIST SP800-22 with a significance level of 0.01.

\begin{table}[!htbp]
\footnotesize 
\caption{NIST SP800-22 test results of 100 million-bit sequences collected from the final outputs. To pass this test suite at a significance level of 0.01, two conditions are required for each of the 15 sub-methods: (i) the P-values-total must be higher than $10^{-4}$, and (ii) the proportion must be higher than 96/100.}
\begin{ruledtabular}
\begin{tabular}{lccc}

\multicolumn{1}{l}{\textbf{Method}}  & \textbf{P-values-total} & \textbf{Proportion} & \textbf{Result} \\ \cmidrule{1-4}
1.   Frequency                  & 0.224 821       & 100/100    & Pass   \\
2.   Block Frequency            & 0.779 188       & 99/100     & Pass   \\
3. Runs                         & 0.096 578       & 99/100     & Pass   \\
4.   Longest Run                & 0.304 126       & 100/100    & Pass   \\
5. Rank                         & 0.574 903       & 99/100     & Pass   \\
6. Fast   Fourier Transform     & 0.366 918       & 98/100     & Pass   \\
7.   Overlapping Template       & 0.003 201       & 99/100     & Pass   \\
8.   Universal                  & 0.911 413       & 100/100    & Pass   \\
9.   Linear Complexity          & 0.015 598       & 100/100    & Pass   \\
10.   Approximate Entropy       & 0.455 937       & 99/100     & Pass   \\
11.   Non-overlapping Template  & 0.514 124       & 100/100    & Pass   \\
12.   Serial                    & 0.249 284       & 99/100     & Pass   \\
13.   Cumulative Sums           & 0.181 557       & 100/100    & Pass   \\
14.   Random Excursions         & 0.062 821       & 99/100     & Pass   \\
15.   Random Excursions Variant & 0.008 266       & 99/100     & Pass   \\ 

\end{tabular}
\end{ruledtabular}
\end{table}

We also used the same dataset to check for the presence of repeating pattern and evaluate the degree of similarity by performing autocorrelation analysis. As plotted in Figure 6, the result indicates no sign of periodic pattern over 100 successive bit intervals. Therefore, the final outputs have no repeating pattern over an acceptable range.

\begin{figure}[!h]
  \includegraphics[width=8cm,keepaspectratio]{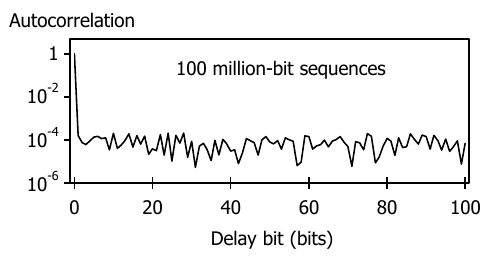}
  \vspace{-2mm}
  \caption{Autocorrelation calculated from a dataset of 100 million-bit sequences collected from the final outputs.}
\end{figure}


To conclude, we investigated the I-V characteristic of a tunnel diode and examined the behavior of its hysteresis cycle. Forward current sweeping takes a slightly different path in each cycle. This random variation is associated with the measured values of $I_p$, and it can serve as an entropy source. We demonstrated a utilization of a time-counting module to measure the values of $t_p$ that are associated with events in which voltages jump to the other branch of the hysteresis loop. As a result, a random distribution was obtained. We demonstrated this proposed technique with inexpensive electronics. Our preliminary prototype used a computer to assess the entropy rate and process the bit-extraction. By harnessing a time-counting module that allowed us to collect signals at a rate of 20 kHz, we generated random bits, featuring stable and almost full entropy rate, at a speed of 90 kbit/s. The final outputs passed the NIST SP800-22\cite{Rukhin}, and autocorrelation was not detected. 

To construct a random number generator with provable robustness against a security attack, implementation of health-monitoring methodologies in accordance to the NIST SP800-90B\cite{Turan} or AIS 20/31\cite{Killmann} standard is mandatory. A proper engineering\cite{Balasch} to detect failure of the hardware to meet industrial certification would be required before real-world use. The work currently in progress includes improvement of the electronic circuit to eliminate environment bias as well as incorporation of a suitable micro-controller or field programmable gate array (FPGA) to optimize the process of data acquisition and post-processing. This development promises the capability to generate random bits, possibly up to 10 Mbit/s.

\section*{Acknowledgements}
This work was supported by a research program from Thailand's National Electronics and Computer Technology Center, the National Science and Technology Development Agency (NECTEC-NSTDA).


\section*{Supplementary Material}
A tunnel diode is a solid-state device and one of many types of semi-conductor diodes incorporated with a p-n junction. In a device, the p-type and n-type are heavily doped. Thus, the tunnel diode is unique in that it exhibits an extremely thin depletion region. Zener diode is another p-n junction diode that is also heavily doped, but Zener diode is incomparable to the tunnel diode as the former has a relatively lower doping concentration.

\setcounter{figure}{0}
\renewcommand{\thefigure}{S\arabic{figure}}

\begin{figure}[!htbp]
\includegraphics[width=5.5cm,keepaspectratio]{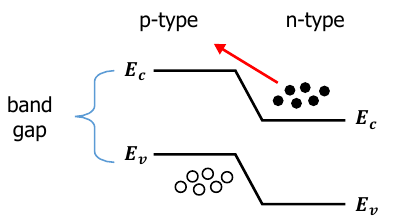}
\caption{Energy level of a p-n semiconductor. The band gap between the conduction band $(E_c)$ and the valence band $(E_v)$ of a typical p-n junction diode is wide and behaves like a thick barrier. This barrier blocks the passage of electrons flowing across the p-n junction. Electrons with energy greater than the barrier can overcome the internal potential force and cause a current flow.  }
\end{figure}

In the case of a typical p-n junction diode, the n-type conduction band aligns at the p-type bandgap (Figure S1). In this scenario, the depletion region behaves like a potential barrier that blocks the passage of electrons flowing from p-type to n-type. The current can flow only when the electrons have energy that surpasses the obstructive force as illustrated in Figure S2a. 

\begin{figure}
\includegraphics[width=8cm,keepaspectratio]{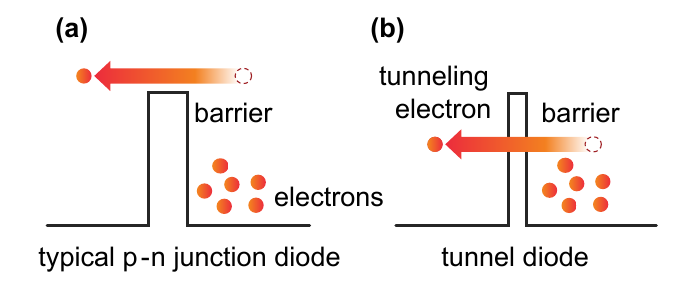}
\caption{Current flow \textbf{(a)} in a typical p-n junction diode where only electrons having energy greater than the potential barrier can cross the barrier and \textbf{(b)} in a tunnel diode having a very thin barrier through which electrons can tunnel.}
\end{figure}

\begin{figure}
\includegraphics[width=7.5cm,keepaspectratio]{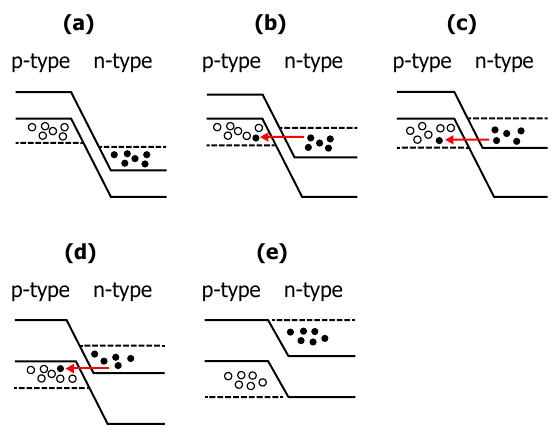}
\caption{Energy levels of the tunnel diode at different scenarios: \textbf{(a)} at normal stage without bias voltage, \textbf{(b)} with just enough bias voltage to cause the conduction band of the n-type to overlap with the valence band of the p-type, \textbf{(c)} when the valence band of the n-type fully overlaps with the valance band of the p-type and current flow is maximally dominated from the quantum tunneling effect, \textbf{(d)} when the overlap grows misaligned with a higher bias voltage such that quantum tunneling effect declines, and \textbf{(e)} at the final stage where high voltage drives the tunnel diode to operate in the typical p-n junction diode.}
\end{figure}

In contrast, a tunnel diode in which electrons can pass through a narrow barrier is depicted in Figure S2b. The device itself can operate in two modes: (i) tunneling mode and (ii) typical p-n junction mode as shown in Figure S3. It can be explained as follows. At a stage where bias voltage drops to zero (Figure S3a), the conduction band of the n-type sits below the valence band of the p-type. By applying small bias (Figure S3b), electrons can tunnel through the depletion region, resulting in the current flow from the conduction band of the n-side to the valence band of the p-side. In this scenario, the rise of the bias voltage increases the probability of electrons flowing, corresponding to the tunneling effect, until the conduction band of the n-side and the valence band of the p-side become fully overlapped (Figure S3c). As a result, there is a peak current $(I_p)$ at a specific bias voltage $(V_p)$. Increasing the voltage level causes these bands to begin to separate (Figure S3d), and the chance of electron tunneling declines until the bands are fully out of alignment at a certain voltage $(V_v)$ (Figure S3e). At $V_v$, the tunneling effect diminishes and therefore provides no contribution to the current flow $(I_v)$. During $V_p$ to $V_v$, where current and biased voltage translate in the opposite direction, negative resistance behavior is exhibited. Applying more voltage will drive the tunnel diode to operate in a typical p-n junction state where current and biased voltage translate in the same direction.

\section*{References and Notes}\bibliography{mybib} 


\end{document}